\documentclass[conference, 10pt]{IEEEtran}
\IEEEoverridecommandlockouts

\usepackage{latexsym}
\usepackage{graphicx}
\usepackage{subcaption}
\usepackage{amsfonts,amssymb,amsmath}
\usepackage{hyperref}
\usepackage{flushend}
\usepackage{cite}
\usepackage{textcomp}
\usepackage{gensymb}
\usepackage{siunitx}
\usepackage{xcolor}
\usepackage{tikz}
\usepackage{etoolbox}
\usepackage{enumitem}
\usepackage{esvect}
\usepackage{mathtools}
\usetikzlibrary{chains,arrows,calc,positioning}
\usepackage{multirow}
\usepackage{comment}
\usepackage[normalem]{ulem}
\usepackage{algorithm}
\usepackage{algorithmic}

\newtoggle{conference}
\toggletrue{conference}  

\def\BibTeX{{\rm B\kern-.05em{\sc i\kern-.025em b}\kern-.08em T\kern-.1667em\lower.7ex\hbox{E}\kern-.125emX}}

\def\nb0{{\mathbf{0}}}
\def\nb1{{\mathbf{1}}}

\def\N{\sigma^2}

\newcommand{\bsym}{\boldsymbol}

\newcommand{\subsf}{\sf \scriptscriptstyle}

\usepackage[acronyms,nonumberlist,nopostdot,nomain,nogroupskip]{glossaries}
\newacronym{quic}{QUIC}{Quick UDP Internet Connections}
\newacronym{3gpp}{3GPP}{3rd Generation Partnership Project}
\newacronym{adc}{ADC}{Analog to Digital Converter}
\newacronym{dac}{DAC}{Digital to Analog Converter}
\newacronym{if}{IF}{Intermediate Frequency}
\newacronym{sma}{SMA}{SubMiniature Version A}
\newacronym{cot}{CoT}{Commercial Off-the-Shelf}
\newacronym{aod}{AoD}{Angle of Departure}
\newacronym{lo}{LO}{local oscillator }

\newacronym{5g}{5G}{5th generation}

\newacronym{aimd}{AIMD}{Additive Increase Multiplicative Decrease}
\newacronym{am}{AM}{Acknowledged Mode}
\newacronym{amc}{AMC}{Adaptive Modulation and Coding}
\newacronym{aqm}{AQM}{Active Queue Management}
\newacronym{awgn}{AGWN}{Additive White Gaussian Noise}
\newacronym{afd}{AFD}{Austin Fire Department}
\newacronym{balia}{BALIA}{Balanced Link Adaptation}
\newacronym{bdp}{BDP}{Bandwidth-Delay Product}
\newacronym{bf}{BF}{Beamforming}
\newacronym{cc}{CC}{Congestion Control}
\newacronym{cdf}{CDF}{Cumulative Distribution Function}
\newacronym{cn}{CN}{Core Network}
\newacronym{cqi}{CQI}{Channel Quality Information}
\newacronym{cp}{CP}{Control Plane}
\newacronym{csirs}{CSI-RS}{Channel State Information - Reference Signal}
\newacronym{dc}{DC}{Dual Connectivity}
\newacronym{dce}{DCE}{Direct Code Execution}
\newacronym{dci}{DCI}{Downlink Control Information}
\newacronym{dl}{DL}{Downlink}
\newacronym{dmr}{DMR}{Deadline Miss Ratio}
\newacronym{dmrs}{DMRS}{DeModulation Reference Signal}
\newacronym{e2e}{E2E}{End-to-End}
\newacronym{ecn}{ECN}{Explicit Congestion Notification}
\newacronym{edf}{EDF}{Earliest Deadline First}
\newacronym{enb}{eNB}{evolved Node Base}
\newacronym{epc}{EPC}{Evolved Packet Core}
\newacronym{es}{ES}{Edge Server}
\newacronym{fdma}{FDMA}{Frequency Division Multiple Access}
\newacronym{fdd}{FDD}{Frequency Division Duplexing}
\newacronym[firstplural=Radio Access Technologies (RATs)]{rat}{RAT}{Radio Access Technology}
\newacronym{fs}{FS}{Fast Switching}
\newacronym{ftp}{FTP}{File Transfer Protocol}
\newacronym{gnb}{gNB}{Next Generation Node Base}
\newacronym{harq}{HARQ}{Hybrid Automatic Repeat reQuest}
\newacronym{hetnet}{HetNet}{Heterogeneous Network}
\newacronym{hh}{HH}{Hard Handover}
\newacronym{hol}{HOL}{Head-of-Line}
\newacronym{ia}{IA}{Initial Access}
\newacronym{imt}{IMT}{International Mobile Telecommunication}
\newacronym{iot}{IoT}{Internet of Things}
\newacronym{los}{LOS}{Line of Sight}
\newacronym{lte}{LTE}{Long Term Evolution}
\newacronym{m2m}{M2M}{Machine to Machine}
\newacronym{mac}{MAC}{Medium Access Control}
\newacronym{mc}{MC}{Multi-Connectivity}
\newacronym{mcs}{MCS}{Modulation and Coding Scheme}
\newacronym{mec}{MEC}{Mobile Edge Cloud}
\newacronym{mi}{MI}{Mutual Information}
\newacronym{mimo}{MIMO}{Multiple Input Multiple Output}
\newacronym{mmwave}{mmWave}{millimeter wave}
\newacronym{mr}{MR}{Maximum Rate}
\newacronym{mss}{MSS}{Maximum Segment Size}
\newacronym{mtd}{MTD}{Machine-Type Device}
\newacronym{mtu}{MTU}{Maximum Transmission Unit}
\newacronym{nfv}{NFV}{Network Function Virtualization}
\newacronym{nlos}{NLOS}{Non Line of Sight}
\newacronym{nr}{NR}{New Radio}
\newacronym{ofdm}{OFDM}{Orthogonal Frequency Division Multiplexing}
\newacronym{pdcch}{PDCCH}{Physical Downlonk Control Channel}
\newacronym{pdcp}{PDCP}{Packet Data Convergence Protocol}
\newacronym{pdsch}{PDSCH}{Physical Downlink Shared Channel}
\newacronym{pdu}{PDU}{Packet Data Unit}
\newacronym{pf}{PF}{Proportional Fair}
\newacronym{pgw}{PGW}{Packet Gateway}
\newacronym{phy}{PHY}{Physical}
\newacronym{pbch}{PBCH}{Physical Broadcast Channel}
\newacronym[plural=\gls{mme}s,firstplural=Mobility Management Entities (MMEs)]{mme}{MME}{Mobility Management Entity}
\newacronym{prb}{PRB}{Physical Resource Block}
\newacronym{pss}{PSS}{Primary Synchronization Signal}
\newacronym{pucch}{PUCCH}{Physical Uplink Control Channel}
\newacronym{pusch}{PUSCH}{Physical Uplink Shared Channel}
\newacronym{rach}{RACH}{Random Access Channel}
\newacronym{ran}{RAN}{Radio Access Network}
\newacronym{red}{RED}{Robotics Emergency Deployment}
\newacronym{rf}{RF}{Radio Frequency}
\newacronym{rlc}{RLC}{Radio Link Control}
\newacronym{rlf}{RLF}{Radio Link Failure}
\newacronym{rrc}{RRC}{Radio Resource Control}
\newacronym{rrm}{RRM}{Radio Resource Management}
\newacronym{rr}{RR}{Round Robin}
\newacronym{rs}{RS}{Remote Server}
\newacronym{rsrp}{RSRP}{Reference Signal Received Power}
\newacronym{rss}{RSS}{Received Signal Strength}
\newacronym{rtt}{RTT}{Round Trip Time}
\newacronym{rw}{RW}{Receive Window}
\newacronym{rx}{RX}{Receiver}
\newacronym{sa}{SA}{standalone}
\newacronym{sack}{SACK}{Selective Acknowledgment}
\newacronym{sap}{SAP}{Service Access Point}
\newacronym{sch}{SCH}{Secondary Cell Handover}
\newacronym{scoot}{SCOOT}{Split Cycle Offset Optimization Technique}
\newacronym{sdma}{SDMA}{Spatial Division Multiple Access}
\newacronym{sinr}{SINR}{Signal to Interference plus Noise Ratio}
\newacronym{sm}{SM}{Saturation Mode}
\newacronym{snr}{SNR}{Signal to Noise Ratio}
\newacronym{son}{SON}{Self-Organizing Network}
\newacronym{ss}{SS}{Synchronization Signal}
\newacronym{srs}{SRS}{Sounding Reference Signal}
\newacronym{sss}{SSS}{Secondary Synchronization Signal}
\newacronym{tb}{TB}{Transport Block}
\newacronym{tcp}{TCP}{Transmission Control Protocol}
\newacronym{tdd}{TDD}{Time Division Duplexing}
\newacronym{tdma}{TDMA}{Time Division Multiple Access}
\newacronym{tfl}{TfL}{Transport for London}
\newacronym{tm}{TM}{Transparent Mode}
\newacronym{trp}{TRP}{Transmitter Receiver Pair}
\newacronym{tti}{TTI}{Transmission Time Interval}
\newacronym{ttt}{TTT}{Time-to-Trigger}
\newacronym{tx}{TX}{Transmitter}
\newacronym{ue}{UE}{User Equipment}
\newacronym{ul}{UL}{Uplink}
\newacronym{uml}{UML}{Unified Modeling Language}
\newacronym{um}{UM}{Unacknowledged Mode}
\newacronym{utc}{UTC}{Urban Traffic Control}
\newacronym{vm}{VM}{Virtual Machine}
\newacronym{rsrq}{RSRQ}{Reference Signal Received Quality}
\newacronym{rssi}{RSSI}{Received Signal Strength Indicator}
\newacronym{crs}{CRS}{Cell Reference Signal}
\newacronym{comp}{CoMP}{Coordinated Multi-Point}
\newacronym{cran}{C-RAN}{Cloud \acrlong{ran}}
\newacronym{ca}{CA}{Carrier Aggregation}
\newacronym{cco}{CC}{Carrier Component}
\newacronym{nsa}{NSA}{Non Stand Alone}
\newacronym{embb}{eMBB}{Enhanced Mobility Broadband}
\newacronym{bsr}{BSR}{Buffer Status Report}
\newacronym{srb}{SRB}{Service Radio Bearer}
\newacronym{scm}{SCM}{Spatial Channel Model}
\newacronym{sctp}{SCTP}{Stream Control Transmission Protocol}
\newacronym{mptcp}{MPTCP}{Multi-path TCP}
\newacronym{ietf}{IETF}{Internet Engineering Task Force}
\newacronym{os}{OS}{Operating System}
\newacronym{tls}{TLS}{Transport Layer Security}
\newacronym{rfc}{RFC}{Request for Comments}
\newacronym{http}{HTTP}{HyperText Transfer Protocol}
\newacronym{nat}{NAT}{Network Address Translation}
\newacronym{api}{API}{Application Programming Interface}
\newacronym{rto}{RTO}{Retransmission Timeout}
\newacronym{psc}{PSC}{Public Safety Communication}
\newacronym{rpgm}{RPGM}{Reference Point Group Mobility}
\newacronym{ic}{IC}{Incident Command}
\newacronym{rsu}{RSU}{Road Side Unit}
\newacronym{uav}{UAV}{unmanned aerial vehicle}
\newacronym{usv}{USV}{Unmanned Surface Vehicle}
\newacronym{uas}{UAS}{Unmanned Aerial System}
\newacronym{iab}{IAB}{Integrated Access and Backhaul}
\newacronym{qoe}{QoE}{Quality of Experience}
\newacronym{ssim}{SSIM}{Structural Similarity Index}
\newacronym{psnr}{PSNR}{Peak Signal to Noise Ratio}
\newacronym{bs}{BS}{Base Station}
\newacronym{mu}{MU}{Multiple User}
\newacronym{ag}{AG}{Air-to-Ground}
\newacronym{af}{AF}{Array Factor}
\newacronym{ula}{ULA}{Uniform Linear Array}
\newacronym{upa}{UPA}{Uniform Planar Array}
\newacronym{lcs}{LCS}{Local Coordinate System}
\newacronym{psd}{PSD}{Power Spectral Density}
\newacronym{vq}{VQ}{vector quantization}
\newacronym{a2g}{A2G}{air-to-ground}
\newacronym{em}{EM}{electromagnetic}
\newacronym{vae}{VAE}{variational autoencoder}

\def\bb0{{\mathbb{0}}}

\def\bb{{\boldsymbol{b}}}

\def\b0{{\boldsymbol{0}}}

\def\b{{\mathrm{b}}}

\def\r0{{\mathbf{0}}}

\def\bsf0{{\bm{\mathsf{0}}}}

\def\N0{{N_{\mathrm{0}}}}

\def\bsf{{\boldsymbol{s}_\mathrm{f}}}

\newcommand{\be}{\begin{equation}}
\newcommand{\ee}{\end{equation}}
\newcommand{\bal}{\begin{align}}
\newcommand{\eal}{\end{align}}

\begin{document}

\title{Near-Field Measurement System \\for the Upper Mid-Band}
\author{
 \IEEEauthorblockN{Ali Rasteh\IEEEauthorrefmark{1},
 Raghavendra Palayam Hari\IEEEauthorrefmark{1},
 Hao Guo\IEEEauthorrefmark{1}\IEEEauthorrefmark{2},
 Marco Mezzavilla\IEEEauthorrefmark{3},
 Sundeep Rangan\IEEEauthorrefmark{1}}
 
\IEEEauthorblockA{\IEEEauthorrefmark{1}Tandon School of Engineering, New York University, Brooklyn, NY, USA\\
\IEEEauthorrefmark{2}Department of Electrical Engineering, Chalmers University of Technology, Gothenburg, Sweden\\
\IEEEauthorrefmark{3}Dipartimento di Elettronica, Informazione e Bioingegneria (DEIB), Politecnico di Milano, Milan, Italy\\
}

\thanks{This work has been supported, in part, by the Swedish Research Council (VR) grant 2023-00272.}
}

\maketitle

\begin{abstract}
The upper mid-band (or FR3, spanning 6-24 GHz) is a crucial frequency range for next-generation mobile networks, offering a favorable balance between coverage and spectrum efficiency. From another perspective, the systems operating in the near-field in both indoor environment and outdoor environments can support line-of-sight multiple input multiple output (MIMO) communications and be beneficial from the FR3 bands. In this paper, a novel method is proposed to measure the 
near-field parameters leveraging a recently developed reflection model where the near-field paths can be described by their image points. We show that these image points
can be accurately estimated via triangulation from multiple measurements with a small number of antennas in each measurement, thus affording a low-cost procedure for near-field multi-path parameter extraction.  A preliminary
experimental apparatus is presented comprising 2 transmit and 2 receive antennas mounted on a linear track to measure the $2 \times 2$ MIMO channel at various displacements.  The system uses a recently-developed wideband radio frequency (RF) transceiver board with fast frequency switching, an FPGA for fast baseband processing, and a new parameter
extraction method to recover paths and spherical characteristics from the multiple $2 \times 2$
measurements.

\end{abstract}


\IEEEpeerreviewmaketitle


\section{Introduction}
\label{sec:intro}

The upper mid-band from 6-24 GHz has attracted considerable attention for new cellular and wireless communications as well as sensing and localization \cite{kang2024cellular,shakya2024angular,shakya2024propagation,park2024end}.   Due to the high frequencies, 
these systems may operate in the near-field to support so-called line-of-sight (LOS) MIMO
systems \cite{9903389,9738442}.  The near-field occurs when the transmit-receive (TX-RX) distance $R$ is below
the Rayleigh distance $R_D$
\begin{align}
    R < R_D = \frac{2D^2}{\lambda},
\end{align}
where $D$ is the aperture of the TX and RX arrays and $\lambda$ is the wavelength.
Accurate channel modeling in the near-field requires capturing the spherical nature of each path.

This paper addresses several critical challenges
facing current channel measurement methods
when applied to learning near-field channels:
\begin{itemize}
    \item \emph{Lack of near-field parameters}:  Typical channel models,
    such as those used by 3GPP \cite{3gpp38901},
    describe each path via its angles of arrival and departure, phase and gain.
    These are the parameters measured by most channel sounders.
    While this model is sufficient for extrapolating the MIMO matrix in the far-field, these path parameters do not describe the spherical nature of propagation necessary for accurate modeling in the near-field.  

    \item \emph{Need for high-dimensional arrays}:  The most straightforward method
    to estimate the channel over a wide aperture
    and capture the spherical wavefronts
    is to use a high-dimensional array with
    a large number of elements spread over a wide range.  Such an experimental system can be costly.

    \item \emph{Multipath}:  The focus of 
    many near-field systems have been in 
    LOS settings.
    However, in many practical settings,
    channels have both LOS and NLOS components,
    and the spherical nature of each
    path must be obtained.   
\end{itemize}

To address these challenges, we present
a novel channel estimation method based on several key contributions:
\begin{itemize}
\item \emph{Use of reflection model
for near-field propagation}:
We extend our recently-developed
\emph{reflection model} (RM) \cite{hu2023parametrization} to describe a multi-path 
channel where any number of the paths 
may be in the near-field.  
While various parametrizations have been proposed for spherical models 
with validation via ray tracing
\cite{10416965,9940939}, the reflection model
enables a simple geometric interpretation and can be applied in the NLOS settings.  

\item \emph{Synthetic aperture from pairwise
non-coherent measurements}:
We provide a procedure to measure the 
near-field channel with only two TX and
two RX elements.  The array elements at both
the TX and RX are moved mechanically, and a synthetic aperture is constructed from the multiple measurements. Importantly, the measurements can be non-coherent allowing significant time between measurements without need for long-term 
phase coherence.

\item \emph{Preliminary experimental demonstration}:  We provide a preliminary 
demonstration of the experimental procedure
leveraging a $2 \times 2$ software defined
radio (SDR) developed by Pi-Radio \cite{mezzavilla2024frequency} for the upper mid-band.  The SDR is mounted on linear tracks
for the synthetic aperture.
\end{itemize}

\noindent
\paragraph*{Other prior work}
The work \cite{10416965} presented a measurement system
for the near-field measurements using VNAs to determine regions in which certain spherical models hold.  This method used a single
antenna mechanically rotated on a turntable  
with the goal of validating ray tracing.
The work \cite{bodet2024sub} measured
and validated models for 
antenna patterns in the sub-THz regime
very close to the antenna.
Both of these works do not consider
the problem of parameter estimation 
in multi-path environments.
A sparse recovery method for parameter estimation is simulated in \cite{yang2024near},
but this method requires a high-dimensional array. Our designed multi-band FR3 channel measurement system has been evaluated in \cite{bomfin2024experimental} however, the near-field aspect was not covered.

\section{Near-Field Channel Model}
\label{sec:nf_model}

\subsection{Overview}
\label{sec:nf_overview}

To capture near-field propagation, we use
the \emph{reflection model} in \cite{hu2023parametrization}.
To simplify the presentation, and to be more consistent with the measurement setup described below,
we consider the special case
of a $p=2$ dimensional model.
That is, all the paths and objects are on a 2D plane. 
However, \cite{hu2023parametrization} provides the mathematical details 
for the $p=3$ dimensional case.  The $p=2$ case for the RM is described as follows:
Let $\bsym{x}^t_0$ and $\bsym{x}^r_0$ be two \emph{reference}
points at the TX and RX.  These reference points would typically
be the centroids in the TX and RX arrays.
We wish to characterize the channel between any two other points:  $\bsym{x}^t$ close to $\bsym{x}^t_0$ and $\bsym{x}^r$ close to $\bsym{x}^r_0$ .
We assume a standard discrete multi-path model where 
the narrowband channel from $\bsym{x}^t$ to $\bsym{x}^r$ is given by
\begin{equation} \label{eq:Hsum_1}
    H(\bsym{x}^r, \bsym{x}^t, f) = \sum_{\ell=1}^L
    g_\ell \exp\left( \frac{2\pi if}{c} d_\ell(\bsym{x}^r, \bsym{x}^t)\right),
\end{equation}
where $f$ is the frequency, 
$L$ is the number of paths, and, for each path $\ell$, $g_\ell$ is its complex path gain, and 
$d_\ell(\bsym{x}^r, \bsym{x}^t)$ is the \emph{path 
distance} between $\bsym{x}^r$  and $\bsym{x}^t$ along the path.

Characterizing the path distance $d_\ell(\bsym{x}^r, \bsym{x}^t)$ as a function of $\bsym{x}^r$ and 
$\bsym{x}^t$ is sufficient to capture the MIMO wideband response for any array.
Specifically, suppose there are 
$N$ TX array elements
at locations $\bsym{x}^t_n$, $n=1,\ldots,N$,
and $M$ RX array elements 
at locations $\bsym{x}^r_m$, $m=1,\ldots,M$.
Then, the MIMO array coefficient is given by:
\begin{equation} \label{eq:Hsum_2}
    H_{mn}(f) = H(\bsym{x}^r_m, \bsym{x}^t_n, f) = \sum_{\ell=1}^L
    g_\ell \exp\left( \frac{2\pi if}{c} d_\ell(\bsym{x}^r_m, \bsym{x}^t_n)\right).
\end{equation}
Hence, if the path distance function 
$d_\ell(\bsym{x}^r, \bsym{x}^t)$ is known
for all paths, the MIMO array response
can be predicted.

If the path index $\ell$ corresponds to a LOS path, the path distance function is simply the Euclidean distance
\begin{equation}
    d_\ell(\bsym{x}^r, \bsym{x}^t) = \| \bsym{x}^r - \bsym{x}^t \|.
\end{equation}
The key in the RM model in \cite{hu2023parametrization} is that the path distance for any NLOS path with specular 
planar reflections is identical to the LOS path distance between the RX point $\bsym{x}^r$ and a \emph{reflected image} of the 
TX point
\begin{equation} \label{eq:drm1}
    d_\ell(\bsym{x}^r, \bsym{x}^t) = \| \bsym{x}^r - \bsym{z}^t_\ell \|,
\end{equation}
where $\bsym{z}^t_\ell$ is the reflection of the TX $\bsym{x}^t$ along path $\ell$.  Following a derivation similar to \cite{hu2023parametrization}, it can
be shown the reflection image of any point $\bsym{x}^t$ is given by
\begin{equation} \label{eq:zt1}
    \bsym{z}^t = \bsym{z}^t_{\ell 0} + s_\ell \bsym{R}(\alpha_\ell ) (\bsym{x}^t-\bsym{x}^t_0).
\end{equation}
where $\bsym{z}^t_0$ is the reflection of the reference point $\bsym{x}^t_0$, $\alpha_\ell$ is a rotation angle
caused by the reflection, $\bsym{R}(\alpha)$ is a rotation matrix
\begin{equation}
    \bsym{R}(\alpha):= \begin{bmatrix}
        \cos(\alpha) & -\sin(\alpha) \\
        \sin(\alpha) &  \cos(\alpha) 
    \end{bmatrix},
\end{equation}
and $s_\ell=\pm 1$ is the reflection parameter with $s_\ell=1$ on an even number of interactions and $s_\ell=-1$ for an odd number of reflections.
The geometric interpretation of \eqref{eq:zt1} is that the image point of any point $\bsym{x}^t$ is $\bsym{x}^t$ rotated by $s_\ell \bsym{R}(\alpha)$ and translated by $\bsym{z}^t_{\ell 0} - s_\ell \bsym{R}(\alpha_\ell) \bsym{x}^t_0 $.  An LOS path corresponds to the special case where
\begin{equation}
    s_\ell = 1, \quad \alpha_\ell = 0, \quad \bsym{z}_{\ell 0}^t = \bsym{x}_0^t.
\end{equation}

\subsection{Angular Representation}
\label{sec:nf_angular}
For the sequel,
it is useful to express the path parameters for the near-field
model in terms of the standard angles of arrival (AoAs)
and angles of departure (AoDs), and time of flight.
To this end, first let: 
\begin{equation}
    \tau_{\ell} := \frac{1}{c}\|\bsym{z}^t_{\ell 0} - \bsym{x}^r_0\|,
\end{equation}
where $c$ is the speed of light so that $\tau_{\ell}$ is the absolute time of flight of the path from the reference point on the RX
$\bsym{x}^r_0$ to the reflected image point $\bsym{z}^t_{\ell 0}$.  Let $\bsym{u}^r_\ell$ be the unit vector in this direction:
\begin{equation} \label{eq:ur}
    \bsym{u}^r_\ell := \frac{1}{c\tau_{\ell}} \left[ \bsym{z}^t_{\ell 0} - \bsym{x}^r_0 \right].
\end{equation}
Since we are in $p=2$ dimensions, we can write 
\begin{equation} \label{eq:urphi}
    \bsym{u}^r_\ell = \bsym{u}(\phi^r_\ell),
\end{equation}
where  $\bsym{u}(\phi)$ is the unit vector with angle $\phi$:
\begin{equation}
    \bsym{u}(\phi) = [\cos(\phi), \sin(\phi)]^\intercal.
\end{equation}
The angle $\phi^r_\ell$ can be interpreted as the angle of arrival (AoA) for path $\ell$.  
Substituting \eqref{eq:ur} into \eqref{eq:drm1} and \eqref{eq:zt1},
we obtain
\begin{equation} \label{eq:drm_rx}
    d_\ell(\bsym{x}^r, \bsym{x}^t) = \| \bsym{x}^r - \bsym{x}^r_0 - c\tau_\ell \bsym{u}_\ell^r - s_\ell R(\alpha_\ell)(\bsym{x}^t-\bsym{x}^t_0) \|.
\end{equation}
We can also write the distance function from the TX,
with some simple algebra:
\begin{equation} \label{eq:drm_tx}
    d_\ell(\bsym{x}^r, \bsym{x}^t) = \| \bsym{x}^t - \bsym{x}^t_0 - c\tau_\ell \bsym{u}_\ell^t - s_\ell R(-\alpha_\ell)(\bsym{x}^r-\bsym{x}^r_0) \|,
\end{equation}
where $\bsym{u}^t_\ell$ is a unit vector in the direction
from the TX reference $\bsym{x}^t_0$ to the reflection of the image of RX.  The vector can be written as
\begin{equation}\label{eq:utphi}
    \bsym{u}^t_\ell = \bsym{u}(\phi^t_\ell),
\end{equation}
where
\begin{equation} \label{eq:alphatr}
    \phi^t_\ell := \phi^r_\ell + \alpha_\ell + \frac{(s+1)\pi}{2},
\end{equation}
represents the AoD of path $\ell$.  

\subsection{Relation to the PWA model}
\label{sec:nf_pwa}
As described in
\cite{hu2023parametrization}, the Plane Wave Approximation (PWA) model
can be seen as a first-order approximation
of the distance function \eqref{eq:drm_rx}.
Specifically, it can be shown that
a first order approximation of $d_\ell(\bsym{x}^r,\bsym{x}^t)$
for $(\bsym{x}^r,\bsym{x}^t)$ close to the reference points
$(\bsym{x}^r_0,\bsym{x}^t_0)$ is given by
\begin{align} \label{eq:dpwa}
\MoveEqLeft
    d_\ell(\bsym{x}^r,\bsym{x}^t) \approx c\tau_0
    \nonumber \\
    & - (\bsym{u}^r_\ell)^\intercal (\bsym{x}^r-\bsym{x}^r_0)
    - (\bsym{u}^t_\ell)^\intercal (\bsym{x}^t-\bsym{x}^t_0)
\end{align}
where $\bsym{u}^t_\ell$ and $\bsym{u}^r_\ell$ are 
the unit vectors 
\eqref{eq:utphi} and \eqref{eq:urphi} along the AoA and AoD
respectively.
Since we are interested in the change of the distance
for points close to the reference, it is not necessary
to know the absolute delay $\tau_\ell$.
Instead, we generally only need the relative delay,
say to the first path, which we will denote as
\begin{equation}
    \delta_\ell = \tau_\ell - \tau_0.
\end{equation}
Along with the complex gain of each path $g_\ell$,
the PWA model thus describes each path with the parameters
\begin{equation} \label{eq:theta_pwa}
    \theta_\ell^{\subsf PWA} := (g_\ell, \delta_\ell, \phi^r_\ell, \phi^t_\ell).
\end{equation}

In contrast, for the RM model we wish to use the exact
path distance function $d_\ell(\bsym{x}^r,\bsym{x}^t)$
in \eqref{eq:drm_rx}, not the first order approximation
in \eqref{eq:dpwa}.  The exact model requires the absolute
time-of-flight $\tau_\ell$ as well as the reflection parity
$s_\ell$ for each path.  Along with the gains,
AoA, and AoD, the parameters for each path are given by
the vector
\begin{equation} \label{eq:theta_rm}
    \theta_\ell^{\subsf RM} := (g_\ell, \tau_\ell, \phi^r_\ell, \phi^t_\ell, s_\ell).
\end{equation}

\section{Proposed Parameter Extraction Method}
\label{sec:param_extract}

We fix the TX and RX reference locations, $\bsym{x}^t_0$ and $\bsym{x}^r_0$ around which we wish to measure the near field
parameters. Our goal is to estimate the number of paths $L$,
and for each path $\ell$, extract the RM parameters
\eqref{eq:theta_rm}.
We proceed in three steps:

\medskip
\noindent
\textbf{1. Synthetic Aperture Measurements:} At both the TX and RX we use arrays with $M$ and $N$ elements each, respectively.
The number of elements $M$ and $N$ may be small.
For example, in the experimental set-up below
$M=N=2$.  An individual measurement
with such a small number of elements
cannot obtain good angular resolution for 
the plane wave parameters, or extract
the near-field parameters.
We thus use multiple measurements
to form a \emph{synthetic wide aperture} by moving the array to widely separated locations.
Specifically, we conduct $K$ measurements
with 
\begin{equation}
 \bsym{x}^t_{kn}, \quad k=1,\ldots,K, \quad
 n=1,\ldots,N,
\end{equation}
denoting the location of the TX element $n$
in measurement $k$;
finally, we let 
\begin{equation}
 \bsym{x}^r_{km}, \quad k=1,\ldots,K, \quad
 m=1,\ldots,M,
\end{equation}
denote the location of the RX element $m$
in measurement $k$.  At each measurement
$k$, we measure the wide-band
complex frequency response. We denote 
$r_{kmn}(f)$ the complex frequency response
at frequency $f$ in measurement $k$
from TX antenna $n$ to RX antenna $m$.

\medskip 
\noindent
\textbf{2. PWA Parameter Extraction:} Next, we  identify the paths and obtain their
PWA parameters.  We assume that there is a
subset of measurements $\mathcal{K}_0 \subset \{1,\ldots,K\}$ such that
the AoA and AoD do not vary significantly,
and the PWA model applies.
In this case,
we know that the complex frequency response should be given by~\cite{heath2018foundations}
\begin{equation}
    r_{kmn}(f) \approx \sum_{\ell=1}^L
    g_{\ell k} a_{km}^r(\phi^r_\ell) a^t_{kn}(\phi_\ell^t)e^{-2\pi f\delta_\ell/c},
\end{equation}
where $L$ is the number of paths; $a_{km}^r(\phi^r)$ is the $m$-th component
of the RX spatial signature in measurement $k$;
$a_{kn}^t(\phi^t)$ is the $n$-th component
of the TX spatial signature in measurement $k$;
and $g_{\ell k}$ is the complex gain
of path $\ell$.  Importantly,
this gain can have an arbitrary phase
since we are not assuming phase coherence
between measurements.
We can then estimate the parameters
by minimizing a MSE cost such as:
\begin{align} \label{eq:pwamin}
    \MoveEqLeft J(\phi^t,\phi^r,\delta) 
        := \sum_{k} \left| 
    \sum_{mnf} r_{kmn}(f)  \right.
    \nonumber \\
    &- \sum_{\ell=1}^L \left.
    g_{\ell k} a_{km}^r(\phi^r_\ell) a_{kn}(\phi_\ell^t)e^{-2\pi f\delta_\ell/c} \right|^2,
\end{align}
where the summation is over a discrete
set of frequencies $f$.
The minimization can be performed
with any sparse solver.
Among various methods, Orthogonal Matching Pursuit (OMP) has emerged as a robust algorithm for this purpose, as it iteratively identifies the most significant components of the sparse channel representation, ensuring efficient reconstruction \cite{tropp2007signal, chen2006theoretical}.

A key feature of the minimization \eqref{eq:pwamin} is that it naturally
exploits different features of different
measurements.  Specifically, for measurements
where the antennas are closely spaced,
we obtain no grating lobes and angular ambiguities.  For measurements that are further apart, we obtain higher resolution.

\medskip
\noindent
\textbf{3. Absolute Time of Flight:} Having obtained  the PWA parameters \eqref{eq:theta_pwa}, the only parameters
needed for the RM model \eqref{eq:theta_rm}
are the absolute time-of-flight $\tau_\ell$
and binary reflection variable $s_\ell = \pm 1$.

Once we have estimated the absolute 
time-of-flight $\tau_\ell$ for one path,
say $\ell=0$, we can estimate the other
times of flight from 
\begin{equation}
    \tau_\ell = \tau_0 + \delta_\ell - \delta_0.
\end{equation}
Hence, we only need to estimate $\tau_\ell$
for one path.

For this purpose, we simply use triangulation.
Recall that, even in an NLOS setting, $\tau_\ell$ is the distance from the RX to 
an image point of the TX.  The image point
will be constant under widely-separated antennas.
Hence, we can obtain two PWA measurements
at two widely separated points.
At each point, we measure the AoA.
Then, we perform any standard triangulation 
method to resolve the distance.

An appealing feature of this approach
is that we only need to perform the triangulation
on the highest SNR path.

\section{Simulations}
\label{sec:simulations}

\begin{figure}[ht]
    \centering
    \begin{subfigure}{\linewidth}
        \centering
        \includegraphics[width=0.9\linewidth]{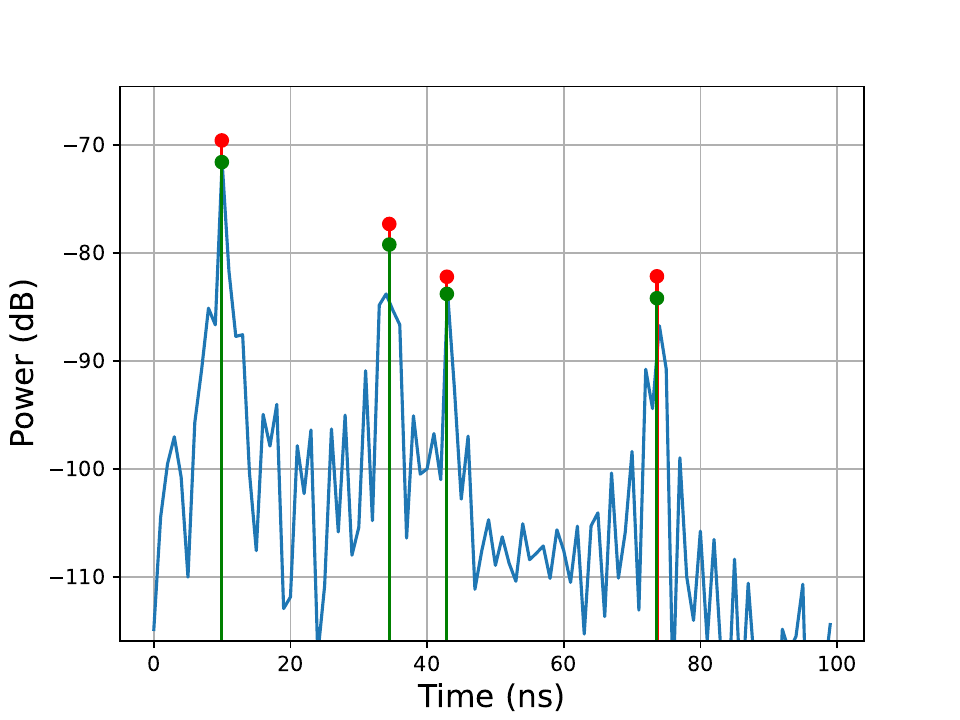}
        \caption{}
        \label{fig:pdp}
    \end{subfigure}


    \begin{subfigure}{\linewidth}
        \centering
        \includegraphics[width=0.9\linewidth]{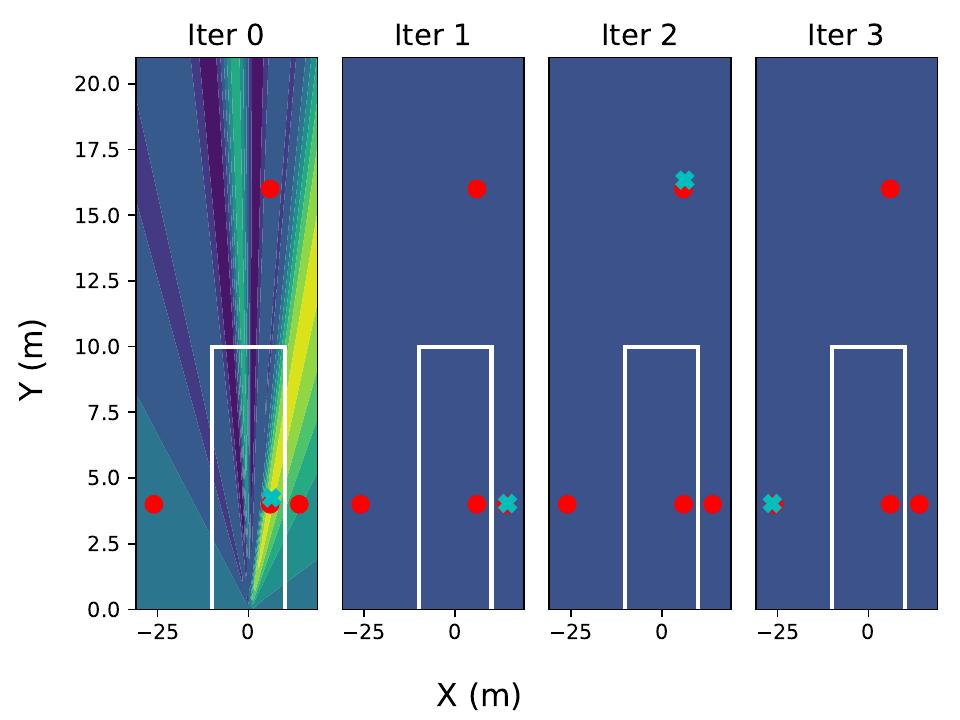}
        \caption{}
        \label{fig:tx_loc}
    \end{subfigure}

    
    \caption{\ref{fig:pdp} Channel Power Delay Profile (PDP) for one transmitter-receiver pair. The red and green dots represent the actual and detected paths, respectively. \ref{fig:tx_loc} A depiction of several iterations in transmitter localization employing the triangulation algorithm is presented. The simulation is carried out in a room measuring 20 meters by 10 meters, utilizing a frequency of 10 GHz and a sampling rate of 1 GHz.}
    \label{fig:simulation}
\end{figure}

A simulation was undertaken to assess the efficacy of the triangulation algorithm. In this experiment, as depicted in Fig.~\ref{fig:simulation}, a 20-meter by 10-meter room is illustrated in white, with a transmitter positioned at the red point within the room. The transmitter is presumed to be isotropic, resulting in three primary reflections from the respective walls, thereby creating three red virtual transmitters outside of the room. Each virtual transmitter constitutes a reflection of the original transmitter relative to the corresponding wall.
Moreover, a receiver equipped with two antennas is initially placed at the origin. Multiple measurements are conducted by configuring the RX antenna with varying antenna spacings ($\lambda/2, \lambda, 2\lambda$) and different positions $\left((0,0), (0.4,0), (0.8,0)\right)$, during which the channel responses are recorded.
In Figure \ref{fig:simulation}, an exemplar channel power delay profile (PDP) for a transmitter-receiver pair is presented, alongside the results obtained with the triangulation algorithm. As illustrated in part (a), the initial step involves estimating the dominant propagation paths from the channel PDP for each specific measurement and transmitter-receiver antenna pair. Subsequently, the locations of the primary transmitter and its corresponding reflection images are determined using triangulation. An illustrative result of the transmitter localization algorithm is demonstrated in Figure \ref{fig:tx_loc} where the estimated locations are represented by blue crosses. The initial iteration illustrates the heatmap depicting the probability of the transmitter's location based on triangulation for the dominant propagation path. Subsequent iterations present the estimated positions of the image points by utilizing information derived from relative delays. It is noteworthy that the performance of the localization algorithm is directly correlated with the SNR of the signals within the simulation.

\section{Experiments}
\label{sec:experiments}

\subsection{Experimental Setup}
\label{sec:setup}

\begin{figure*}[t]
    \centering
    \begin{subfigure}[t]{0.66\textwidth}
        \vspace{0pt}
        \centering
        \includegraphics[width=\textwidth]{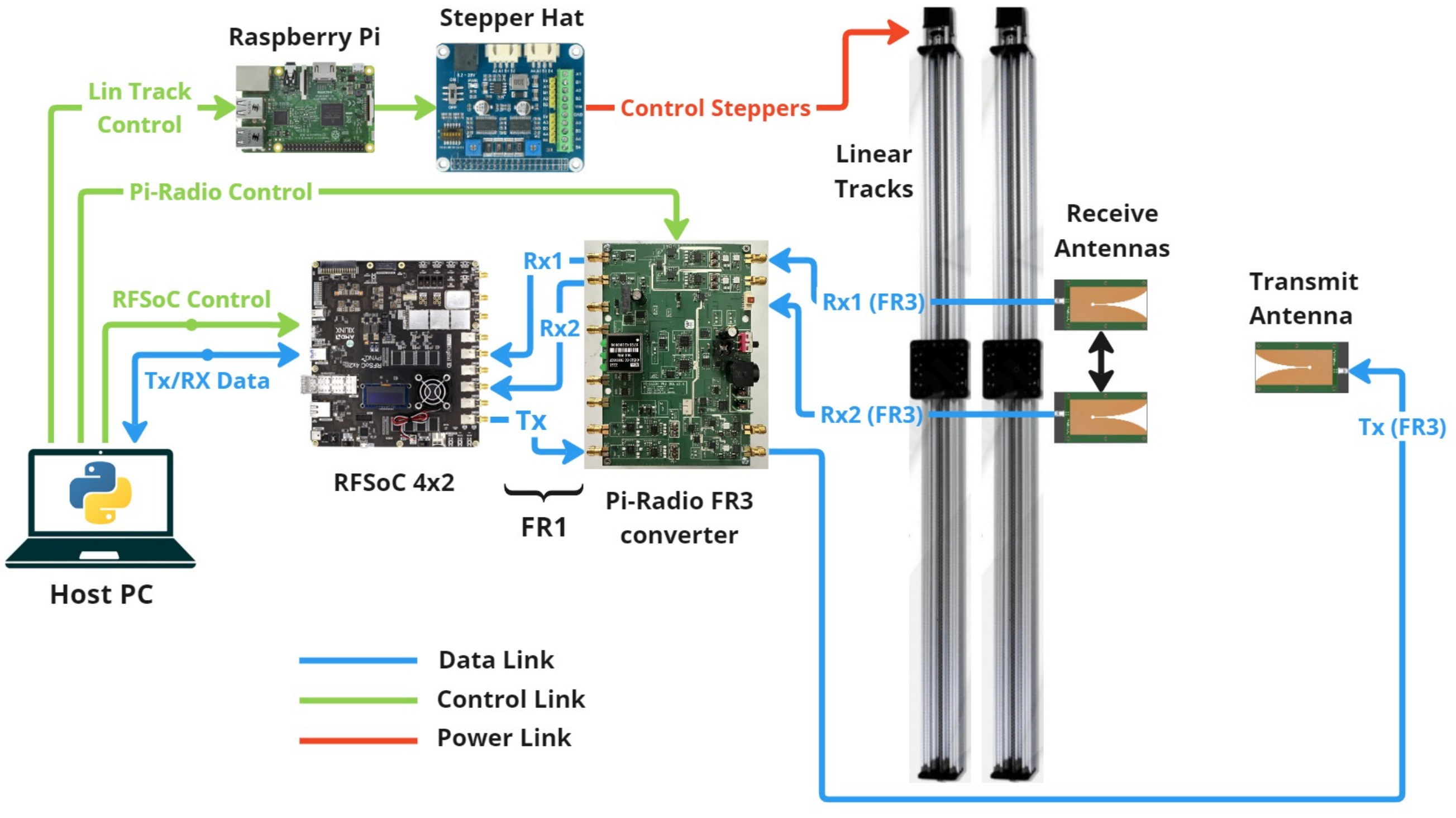}
        \caption{Schematic Diagram of the Experimental Setup}
        \label{fig:setup_schematic}
    \end{subfigure}%
    \hfill
    \begin{subfigure}[t]{0.33\textwidth}
        \vspace{0pt}
        \centering
        \includegraphics[height=0.125\textheight]{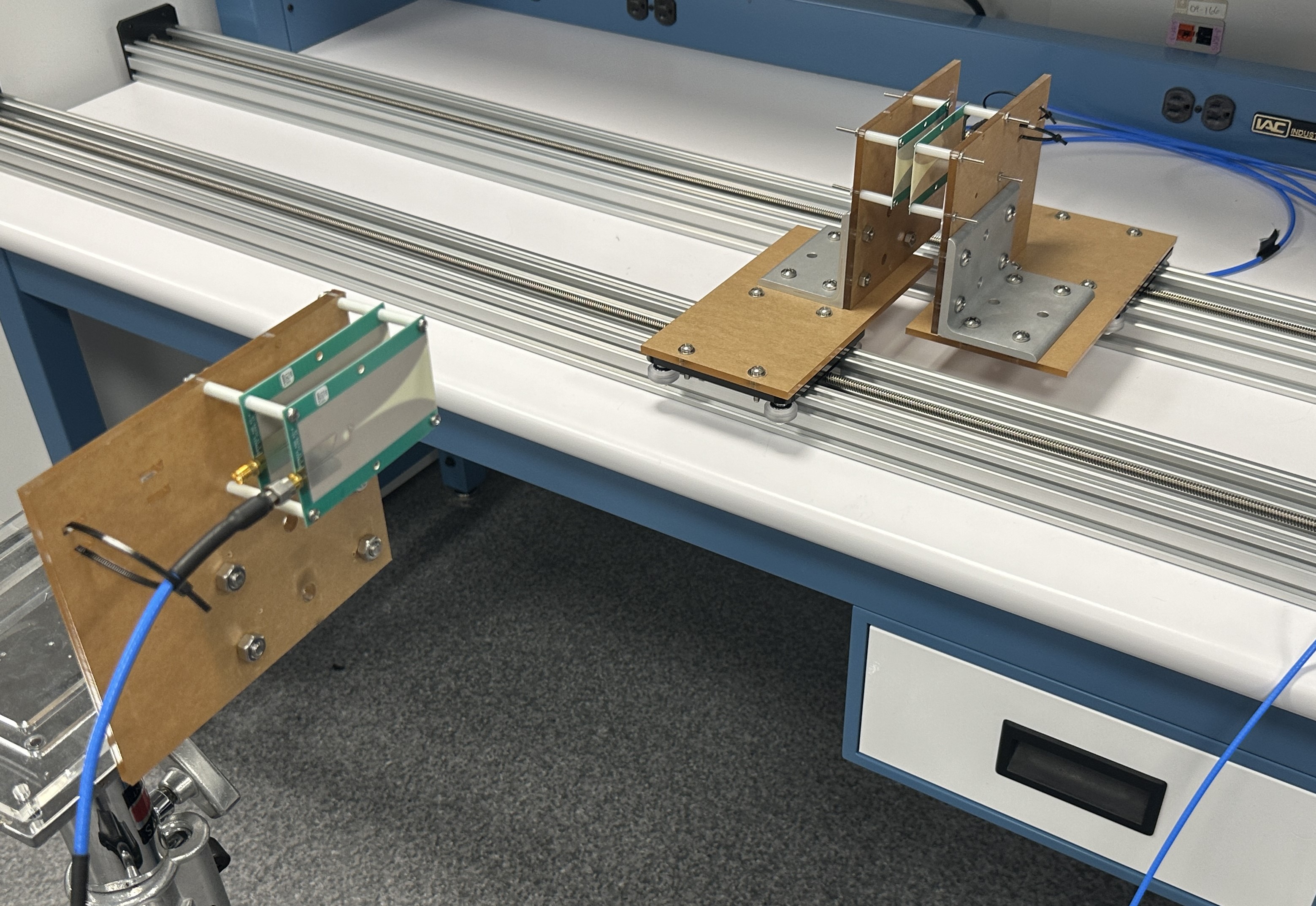}
        \caption{Two linear tracks for wide aperture configuration}
        \label{fig:setup_lintrack}
        \vspace{1em}
        
        \includegraphics[height=0.125\textheight]{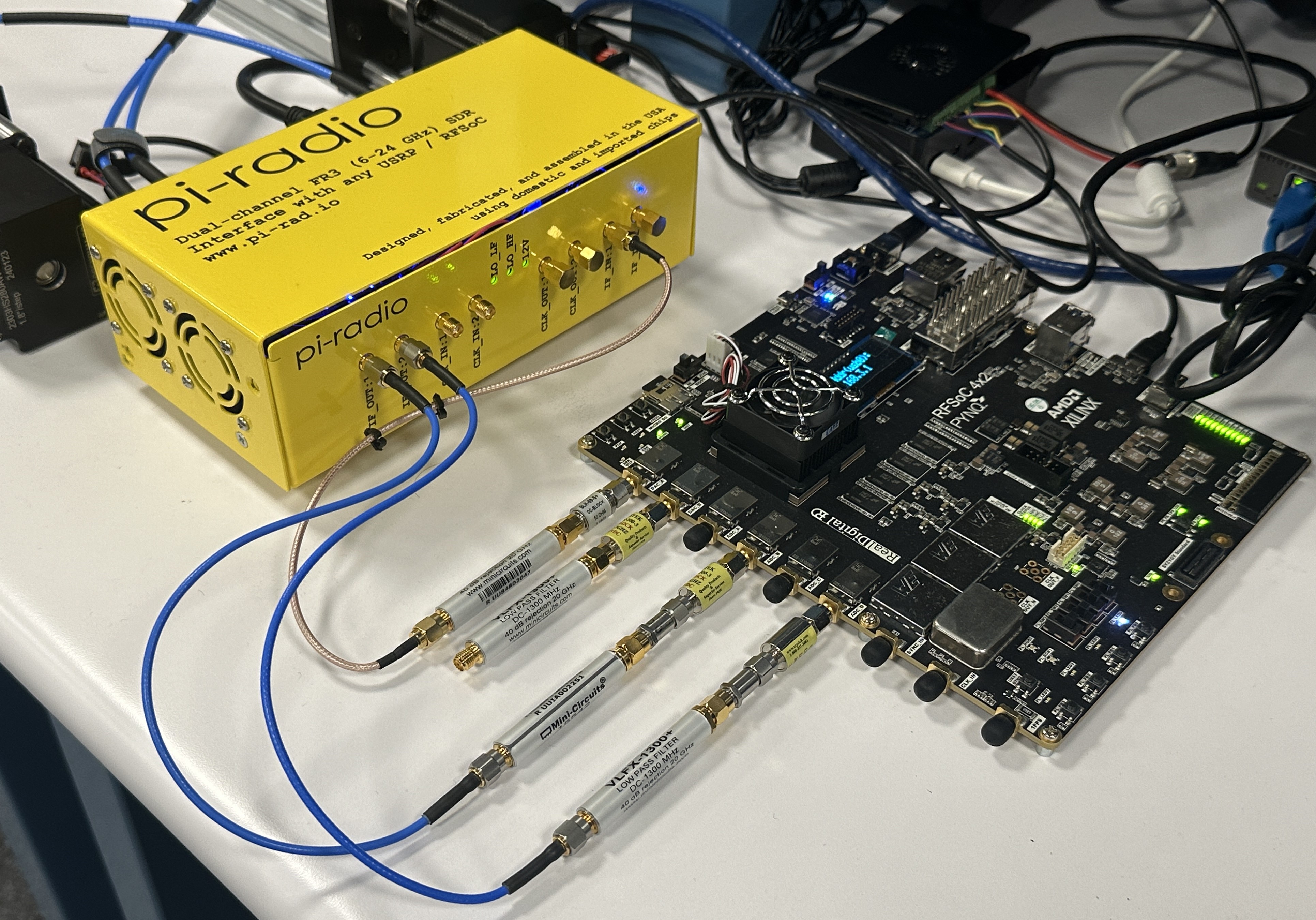}
        \caption{Pi-Radio FR3 transceiver along with the Xilinx RFSoC 4x2}
        \label{fig:setup_transceivers}
    \end{subfigure}
    
    \caption{Our designed FR3 near-field measurement system with Xilinx RFSoC 4x2, Pi-Radio FR3 transceiver, Vivaldi antennas, and two linear tracks to implement the synthetic aperture.}
    \label{fig:setup}
    
\end{figure*}

A preliminary experimental setup is
shown in Fig.~\ref{fig:setup}.
The primary components of the proposed setup are as follows:
\begin{itemize}
    \item \textbf{Xilinx RFSoC:} The Xilinx RFSoC 4x2 serves as the designated intermediate and base-band frequency processor in our experimental setup. This choice is particularly advantageous due to its inclusion of a programmable FPGA and an ARM-based processor, and the support of a wide operating bandwidth.
    \item \textbf{Pi-Radio FR3 RF Transceiver:} This refers to a programmable FR3 Software-Defined Radio (SDR) \cite{mezzavilla2024frequency} developed by Pi-Radio, which is capable of up-converting and down-converting signals between the FR3 and FR1 frequency bands. This transceiver is programmable to operate within 6-24GHz.
    \item \textbf{Vivaldi Antenna:} These antennas are ultra-wideband end-fire types, operable across the entire FR3 frequency range, designed by Pi-Radio. These are utilized for both transmission and reception purposes.
    \item \textbf{Host Computer:} The Host computer functions as a real-time data processing module, responsible for generating transmit frames, retrieving received frames, and subsequently processing and presenting such data in the form of graphs and heat maps in real-time.
    The identical version of a (\href{https://github.com/ali-rasteh/RFSoC_SDR}{Python-based software}) developed at NYU Wireless is deployed on both the host computer and the Xilinx RFSoC 4x2. This software is capable of managing the RFSoC, the Pi-Radio transceiver, the linear tracks' driver, and executing all necessary signal processing functions, including the generation of the transmission (TX) signal, retrieval of the received (RX) signal, and computation of the channel response.
    \item \textbf{Linear Track:} The experimental setup utilizes two controllable linear tracks that function to simulate a synthetic wide aperture antenna array. Each linear track accommodates one receive antenna, which is mounted on movable gantry plates.
    \item \textbf{Raspberry-Pi:} The Raspberry-Pi is employed to regulate and synchronize the movable components on the linear track, thereby determining the positions and spacing of the receive antennas.
    
\end{itemize}

The Host PC serves as the essential cornerstone of the entire configuration, controlling all operational aspects of its component elements. The interface linking the Host PC and the Xilinx RFSoC 4x2 functions as both a control link and a data link, facilitating the transmission and reception of frames. The RFSoC's DAC output produces a 500\si{MHz} wideband signal centered at \si{1GHz}, which is subsequently connected to the intermediate frequency (IF) input of the Pi-Radio transceiver. The Pi-Radio transceiver executes an up-conversion of the IF signal to a 10.0\si{GHz} radio frequency (RF) signal, which is then passed to the Vivaldi transmitter antennas. The inverse process happens in the receive chain, starting with down-conversion in the Pi-Radio transceiver and concluding with the RFSoC 4x2 analog-to-digital converter (ADC).
The transmitting antenna array is installed on a stationary, fixed stand and is oriented toward the receiving antennas. Conversely, the receiving antennas are mounted on the mobile platform of the linear tracks, oriented toward the transmitting antenna. This configuration permits the conceptualization of the receiving antenna arrangement as a wide aperture antenna array as the plates are displaced over time.

\subsection{Experimental Procedure}
\label{sec:exp_procedure}
As elaborated throughout this study, our objective is to estimate the parameters of the reflection model through the employment of the proposed setup. This paper utilizes a single transmitter and two receiver Vivaldi antennas. The receiver antennas are positioned at various offsets and antenna spacings by means of a Raspberry Pi-based controller operating on two linear tracks. In our experiments, three offsets were utilized, 0m, 0.3m, and 0.6m, in conjunction with three distinct antenna spacings $\lambda/2$, $\lambda$, and $2\lambda$, resulting in multiple measurement sets where the channel impulse response was recorded. Subsequently, the line-of-sight (LOS) and reflection image transmitters are identified from the channel responses using Orthogonal Matching Pursuit (OMP) and triangulation as detailed in Sections \ref{sec:param_extract}-\ref{sec:simulations}. Finally, the Plane Wave Approximation (PWA) and Reflection Model (RM) parameters are extracted using the outlined methods.


\section{Conclusion}

We have presented a simple method for extracting
near-field parameters in multi-path
environments.  The method leverages
the reflection model \cite{hu2023parametrization}
that reduces the estimation problem
to a localization problem of image points.
A synthetic aperture procedure can then
be used to measure the necessary parameters
with a small number of non-coherent
measurements.  The method is validated both in a simulation and using a preliminary experimental procedure.  Future work will conduct more extensive simulations and also validate the
experimental measurements in a controlled environment with known true parameters.

\bibliographystyle{IEEEtran}

\begin{thebibliography}{10}
\providecommand{\url}[1]{#1}
\csname url@samestyle\endcsname
\providecommand{\newblock}{\relax}
\providecommand{\bibinfo}[2]{#2}
\providecommand{\BIBentrySTDinterwordspacing}{\spaceskip=0pt\relax}
\providecommand{\BIBentryALTinterwordstretchfactor}{4}
\providecommand{\BIBentryALTinterwordspacing}{\spaceskip=\fontdimen2\font plus
\BIBentryALTinterwordstretchfactor\fontdimen3\font minus \fontdimen4\font\relax}
\providecommand{\BIBforeignlanguage}[2]{{%
\expandafter\ifx\csname l@#1\endcsname\relax
\typeout{** WARNING: IEEEtran.bst: No hyphenation pattern has been}%
\typeout{** loaded for the language `#1'. Using the pattern for}%
\typeout{** the default language instead.}%
\else
\language=\csname l@#1\endcsname
\fi
#2}}
\providecommand{\BIBdecl}{\relax}
\BIBdecl

\bibitem{kang2024cellular}
S.~Kang, M.~Mezzavilla, S.~Rangan, A.~Madanayake, S.~B. Venkatakrishnan, G.~Hellbourg, M.~Ghosh, H.~Rahmani, and A.~Dhananjay, ``Cellular wireless networks in the upper mid-band,'' \emph{IEEE Open J. Commun. Soc.}, Mar. 2024.

\bibitem{shakya2024angular}
D.~Shakya, M.~Ying, and T.~S. Rappaport, ``{Angular Spread Statistics for 6.75 GHz FR1 (C) and 16.95 GHz FR3 Mid-Band Frequencies in an Indoor Hotspot Environment},'' \emph{arXiv preprint arXiv:2409.03013}, 2024.

\bibitem{shakya2024propagation}
D.~Shakya, M.~Ying, T.~S. Rappaport, H.~Poddar, P.~Ma, Y.~Wang, and I.~Al-Wazani, ``{Propagation measurements and channel models in Indoor Environment at 6.75 GHz FR1 (C) and 16.95 GHz FR3 Upper-mid band Spectrum for 5G and 6G},'' \emph{arXiv preprint arXiv:2405.01358}, 2024.

\bibitem{park2024end}
J.~Park, F.~Sohrabi, A.~Ghosh, and J.~G. Andrews, ``{End-to-End Deep Learning for TDD MIMO Systems in the 6G Upper Midbands},'' \emph{arXiv preprint arXiv:2402.01033}, 2024.

\bibitem{9903389}
M.~Cui, Z.~Wu, Y.~Lu, X.~Wei, and L.~Dai, ``{Near-field MIMO communications for 6G: Fundamentals, challenges, potentials, and future directions},'' \emph{IEEE Commun. Mag.}, vol.~61, no.~1, pp. 40--46, Jan. 2023.

\bibitem{9738442}
H.~Zhang, N.~Shlezinger, F.~Guidi, D.~Dardari, M.~F. Imani, and Y.~C. Eldar, ``{Beam focusing for near-field multiuser MIMO communications},'' \emph{IEEE Trans. Wireless Commun.}, vol.~21, no.~9, pp. 7476--7490, Sept. 2022.

\bibitem{3gpp38901}
{3GPP Technical Report 38.901}, ``Study on channel model for frequencies from 0.5 to 100 {GHz} ({R}elease 16),'' Dec. 2019.

\bibitem{hu2023parametrization}
Y.~Hu, M.~Yin, S.~Rangan, and M.~Mezzavilla, ``{Parametrization and estimation of high-rank line-of-sight MIMO channels with reflected paths},'' \emph{IEEE Trans. Wireless Commun.}, Apr. 2024.

\bibitem{10416965}
Z.~Yuan, J.~Zhang, V.~Degli-Esposti, Y.~Zhang, and W.~Fan, ``{Efficient ray-tracing simulation for near-field spatial non-stationary mmWave massive MIMO channel and its experimental validation},'' \emph{IEEE Trans. Wireless Commun.}, pp. 1--1, 2024.

\bibitem{9940939}
Z.~Yuan, J.~Zhang, Y.~Ji, G.~F. Pedersen, and W.~Fan, ``{Spatial non-stationary near-field channel modeling and validation for massive MIMO systems},'' \emph{IEEE Trans. Antennas Propag.}, vol.~71, no.~1, pp. 921--933, Jan. 2023.

\bibitem{mezzavilla2024frequency}
M.~Mezzavilla, A.~Dhananjay, M.~Zappe, and S.~Rangan, ``{A Frequency Hopping Software-Defined Radio Platform for Communications and Sensing in the Upper Mid-Band},'' in \emph{Proc.\ IEEE International Workshop on Signal Processing Advances in Wireless Communications (SPAWC)}, 2024, pp. 611--615.

\bibitem{bodet2024sub}
D.~Bodet, V.~Petrov, S.~Petrushkevich, and J.~M. Jornet, ``{Sub-terahertz near field channel measurements and analysis with beamforming and Bessel beams},'' \emph{Scientific Reports}, vol.~14, no.~1, p. 19675, 2024.

\bibitem{yang2024near}
S.~Yang, Y.~Peng, W.~Lyu, Y.~Li, H.~He, Z.~Zhang, and C.~Yuen, ``{Near-field channel estimation for extremely large-scale Terahertz communications},'' \emph{Science China Information Sciences}, vol.~67, no.~9, p. 192302, 2024.

\bibitem{bomfin2024experimental}
R.~Bomfin, A.~Bazzi, H.~Guo, H.~Lee, M.~Mezzavilla, S.~Rangan, J.~Choi, and M.~Chafii, ``An experimental multi-band channel characterization in the upper mid-band,'' \emph{arXiv preprint arXiv:2411.12888}, 2024.

\bibitem{heath2018foundations}
R.~W. Heath~Jr. and A.~Lozano, \emph{Foundations of {MIMO} Communication}.\hskip 1em plus 0.5em minus 0.4em\relax Cambridge University Press, 2018.

\bibitem{tropp2007signal}
J.~A. Tropp and A.~C. Gilbert, ``Signal recovery from random measurements via orthogonal matching pursuit,'' \emph{IEEE Transactions on information theory}, vol.~53, no.~12, pp. 4655--4666, 2007.

\bibitem{chen2006theoretical}
J.~Chen and X.~Huo, ``Theoretical results on sparse representations of multiple-measurement vectors,'' \emph{IEEE Transactions on Signal processing}, vol.~54, no.~12, pp. 4634--4643, 2006.

\end{thebibliography}

\end{document}